\documentclass[onecolumn,preprintnumbers,amsmath,amssymb,nofootinbib,12pt]{revtex4}
\usepackage{graphicx}
\usepackage{dcolumn}
\usepackage{bm}

\input psfig.sty

\newcommand{\bq}{\begin{equation}}
\newcommand{\eq}{\end{equation}}
\newcommand{\bqn}{\begin{eqnarray}}
\newcommand{\eqn}{\end{eqnarray}}

\newcommand{\lb}{\label}

\def\gappr{\lower 3pt\hbox{$\buildrel > \over \sim\;$}}
\def\gappl{\lower 3pt\hbox{$\buildrel < \over \sim\;$}}
\def\limiter{\lower 7pt\hbox{$\buildrel{\textstyle\longrightarrow}\over{\scriptscriptstyle ~~s\rightarrow\infty~~}\;$}}
\def\ablim{\lower 9pt\hbox{$\buildrel{\textstyle\longrightarrow}\over{\scriptscriptstyle ~~~a\rightarrow b~~~}\;$}}
\def\x0lim{\lower 11pt\hbox{$\buildrel{\textstyle\longrightarrow}\over{\scriptscriptstyle ~~x^0\rightarrow-\infty~~}\;$}}
\def\xlim{\lower 8.5pt\hbox{$\buildrel{\textstyle\longrightarrow}\over{\scriptscriptstyle ~~x_\pm\rightarrow-\infty~~}\;$}}
\def\T0lim{\lower 11pt\hbox{$\buildrel{\textstyle\longrightarrow}\over{\scriptscriptstyle ~~T\rightarrow0~~}\;$}}
\def\Tlim{\lower 8.5pt\hbox{$\buildrel{\textstyle\longrightarrow}\over{\scriptscriptstyle~~T\rightarrow\infty~~}\;$}}
\def\Tmg1{\lower 8.5pt\hbox{$\buildrel{\textstyle\longrightarrow}\over{\scriptscriptstyle~~T>>1~~}\;$}}
\def\pdot{\raise 1.5pt\hbox{.}}

\def\dal{\hbox{$\sqcup$\hbox to 0pt{\hss$\sqcap$}}}

         \def\ln{{\rm ln}}
         \def\ex{{\rm e}}
         \def\dal{\hbox{$\sqcup$\hbox to 0pt{\hss$\sqcap$}}}

\begin{document}
            
\title{\Large SQM studied in the Field Correlator Method}
\author{F. I. M. Pereira}\email{flavio@on.br}
\affiliation{Observat\'orio Nacional, MCTI, Rua Gal. Jos\'e Cristino 77, 
20921-400 Rio de Janeiro RJ, Brazil}
\date{\today}

\begin{abstract}
 By using the recent nonperturbative equation of state of the quark gluon plasma 
derived in the formalism of the Field Correlator Method, we investigate the bulk 
properties of the strange quark matter in beta-equilibrium and with charge neutrality 
at $T=p=0$. 
 The results show that the stability of strange quark matter with respect to $^{56}F_\ex$
is strongly dependent on the model parameters, namely, the gluon condensate $G_2$ and the 
q$\bar{\rm q}$ interaction potential $V_1$. 
 A remarkable result is that the width of the stability window decreases as $V_1$ increases, 
being maximum at $V_1=0$ and nearly zero at $V_1=0.5$ GeV. 
 For $V_1$ in the range $0\leq V_1\leq0.5$ GeV, all values of $G_2$ are lower than 
$0.006-0.007\;{\rm GeV}^4$ obtained from comparison with lattice results at 
$T_c\;(\mu=0)\sim170$ MeV. 
 These results do not favor the possibilities for the existence of (either nonnegative  
or negative) absolutely stable strange quark matter. 
\end{abstract}

\maketitle

 
\section{\bf Introduction}
\lb{int}
  
 Since the works of A. R. Bodmer \cite{Bod} and E. Witten \cite{Wit}, the existence of strange quark 
matter (SQM) has been largely investigated.
 SQM is a particular form of matter comprising roughly equal amounts of $u$, $d$ and $s$ quarks.
 It is presumed that SQM has been produced at extreme conditions of high temperatures and 
densities in the beginning of the universe and/or latter at low temperatures and high 
densities in compact stellar interiors (e. g., neutron stars and quark stars).
 In this regard, it is of importance to mention the pioneer work of N. Itoh, who investigated 
the possibilities for the existence of hypothetical quark stars \cite{Ito}.

 According to the Bodmer-Witten conjecture, SQM might be more stable than ordinary nuclear matter.  
 In a pioneer work, E. Farhi and R. L. Jaffe \cite{Far} investigated SQM in equilibrium with respect to 
weak interactions, at zero temperature and pressure, in the context of the MIT Bag Model \cite{MIT}.
 In this model, quarks enter in the respective equation of state (EOS) as a free quark gas with a 
Fermi-Dirac distribution.
 The confinement is represented by a bag enclosing the free quarks with a constant $B$ 
which gives the vacuum energy density difference between the confined and deconfined phases. 
 Improvement of the model is obtained by the inclusion of corrections to first order in 
the QCD coupling constant in the $\alpha_c<1$ (perturbative) regime (see \cite{Far} and 
references therein). 
 Until now, SQM properties in dense nuclear matter and compact stars interiors have mostly been 
considered in the framework of the MIT Bag Model 
\cite{Far,HZS,AFO,KWW,Mad,DPM1,DPM2,Koh,Alb,Bur1,Bur2,Bur3,Sag,Boe,Pag,Min}.  
 Applications of the Nambu-Jona-Lasinio \cite{NJL1,NJL2} quantum field theoretical approach  have 
also been done to describe quark matter properties in compact stars interiors 
\cite{DPM1,DPM2,BB1,BB2,BB3,LBT,Bla}.
 In alternative investigations, in which quark masses are density dependent, SQM is also taken as a 
free Fermi gas mixture of quarks and anti-quarks \cite{BeL,LuB}.
 These approaches are used to describe $u$, $d$ and $s$ quark matter at zero and (not so high) nonzero 
temperatures and large density regions where the approximation of free quarks can be considered. 
 However, this is not so at all densities and (low) temperatures.
 Quarks strongly interact subjected to a potential that is large when the ${\rm q\bar q}$ distances 
are large. 
 In this case, nonperturbative methods must be considered.

 QCD, the fundamental theory of strong interactions, due to its nonlinearity has been taken 
as an inappropriate theory for the purposes of practical applications as, for instance, 
the calculation of an EOS to describe quark matter at all finite temperatures and densities, 
including nonperturbative effects of confinement. 
 Asymptotically, the description of quark matter becomes simple, but at low temperatures and 
(or moderate) densities the attempts to obtain an EOS including the confinement have appeared 
as a difficult task to be achieved. 
 However, since 1987, the investigations of Yu. A. Simonov \cite{Si1} and H. G. Dosch 
\cite{HGD,DSi,DPS} have resulted in the construction of an important method based on vacuum 
field correlators functions that is being continuously developed up to now.

 Recently the nonperturbative EOS of quark-gluon plasma was derived in the framework of the 
Field Correlator Method (FCM) \cite{Si6}, also called Stochastic Vacuum Model. 
 FCM (for a review see \cite{DiG} and references therein) is a nonperturbative approach which 
naturally includes from first principles the dynamics of confinement in terms of color electric 
and color magnetic correlators. 
 The parameters of the model are the gluon condensate $G_2$ and the ${\rm q\bar q}$ interaction 
potential $V_1$ which govern the behavior of the EOS, at fixed quark masses and temperature.
 FCM has been used to describe the quark-gluon plasma dynamics and phase transition 
\cite{ST1,ST2,KS1,KS2}. 
 An important feature of the model is that it covers the entire phase 
diagram plane, from the large temperature and small density region to the small temperature and 
large density region.
  In the connection between FCM and lattice simulations, the critical temperature at 
$\mu_c=0$ turns out to be $T_c\sim170$ MeV for $G_2\simeq0.006-0.007\;{\rm GeV}^4$ \cite{ST1,ST2}.
 
 In a previous work, the existence of stable SQM on strange star surfaces (see Sec.3.3 in 
\cite{Fla}) was shortly considered in the FCM framework. 
 In the present work, we perform a more detailed investigation of SQM on the same line. 
 The system is a gas of $u$, $d$ and $s$ quarks and gluons subjected to the interaction 
potential $V_1$.
 The vacuum energy density difference between confined and deconfined phases, 
$\Delta|\varepsilon_{vac}|$, is a nonperturbative quantity expressed in terms of $G_2$. 
 The main parameters of our calculation are $\Delta|\varepsilon_{vac}|$(or $G_2$) and $V_1$, 
and the strange quark mass $m_s$ (assuming $m_u=m_d=0$). 
 We first investigate the energy per baryon, from which we obtain the stability window with 
respect to the $^{56}F_\ex$ nucleus.
 Strangeness, hadronic electric charge and density are also addressed.

 This application of the FCM to the study of the bulk properties of SQM (not considered before) 
shows the role of the method to provide alternative indications for its parameters.
 In \cite{Fla} we have shown the importance of the comparison of the FCM calculations with 
astrophysical observations of some strange star candidates. 
 Similarly, in the present work, the main purpose is the relevance of the FCM predictions for 
the SQM properties to be compared with lattice simulations and/or the results at RHIC and LHC 
experiments. 

 This paper is organized as follows. 
 In Sec. \ref{beqs} we summarize the theoretical framework of the FCM and show the equations to 
be used in our calculation.
 In Sec. \ref{res} we show the results and in Sec. \ref{frem} we give the final remarks and 
conclusions. 

\section{Basics equations}
\lb{beqs}

 In the FCM approach, the confined-deconfined phase transition is dominated by the 
nonperturbative correlators \cite{DiG}. 
 The dynamic of deconfinement is described by Gaussian (quadratic in $F^a_{\mu\nu}F^a_{\mu\nu}$)  
colorelectric and colormagnetic gauge invariant  Fields Correlators $D^E(x)$, $D^E_1(x)$, 
$D^H(x)$, and $D^H_1(x)$. 
 The main quantity which governs the nonperturbative dynamics of deconfinement is given by the 
two point functions (after a decomposition is made) 
\bqn
g^2\bigg<\hat tr_f[E_i(x)\Phi(x,y)E_k(x)\Phi(y,x)]\bigg>_B&=&\delta_{ik}
[D^E+D^E_1+u^2_4\frac{\partial D^E_1}{\partial u^2_4}]+
u_iu_k\frac{\partial D^E_1}{\partial u^2}
\lb{gEE}
\eqn
\bqn
g^2\bigg<\hat tr_f[H_i(x)\Phi(x,y)H_k(x)\Phi(y,x)]\bigg>_B&=&\delta_{ik}
[D^H+D^H_1+u^2_4\frac{\partial D^H_1}{\partial u^2_4}]-
u_iu_k\frac{\partial D^H_1}{\partial u^2}
\lb{gHH}
\eqn
where $u=x-y$ and 
\bq
\Phi(x,y)=P\exp\bigg[ig\int^y_xA_\mu dx^\mu\bigg]
\lb{29}
\eq
is the parallel transporter (Schwinger line) to assure gauge invariance.

  In the confined phase (below $T_c$), $D^E(x)$ is responsible for confinement with string 
tension $\sigma^E=(1/2)\int D^E(x)d^2x$. 
 Above $T_c$ (deconfined phase), $D^E(x)$ vanishes while $D^E_1(x)$ remains nonzero being 
responsible (toghether with the magnetic part due to $D^H(x)$ and $D^H_1(x)$) for nonperturbative 
dynamics of the deconfined phase. 
  In lattice calculations, the nonperturbative part of $D^E_1(x)$ is parametrized in the form 
  \cite{Si5,DiG}
\bq
D^E_1(x)=D^E_1(0)\ex^{-|x|/\lambda}\;,
\lb{DE1}
\eq
where $\lambda=0.34\;{\rm fm}$ {\rm(full QCD)} is the correlation length, with the normalization 
fixed at $T=\mu=0$,
\bq
D^E(0)+D^E_1(0)=\frac{\pi^2}{18}G_2\;,
\lb{DEDE1}
\eq
where $G_2$ is the gluon condensate \cite{SVZ}.
 
 The generalization of the FCM at finite $T$ and $\mu$ provides expressions  for the 
thermodynamics quantities where the leading contribution is given by the interaction of 
the single quark and gluon lines with the vacuum (called single line approximation (SLA)).  
 As in \cite{Fla}, from \cite{Si6} and standard thermodynamical relations \cite{Kap}, we here 
explicitly rewrite in more convenient forms (for our purposes) the expressions (for one quark 
system, $N_f=1$) for the pressure 
\bq
p^{SLA}_q=\frac{1}{3}
\frac{2N_c}{(2\pi)^3}\int d^3k\frac{k^2}{E}
\bigg[f^{SLA}_q(T,J^E_1,\mu_q)+{\bar f}^{SLA}_q(T,J^E_1,\mu_q)\bigg]\;,
\lb{p}
\eq
energy density
\bq
\varepsilon^{SLA}_q=
\frac{2N_c}{(2\pi)^3}\int d^3kÅœ
\bigg[E-T(T\frac{\partial J^E_1}{\partial T}+\mu_q\frac{\partial J^E_1}{\partial\mu_q})\bigg]
\bigg[f^{SLA}_q(T,J^E_1,\mu_q)+{\bar f}^{SLA}_q(T,J^E_1,\mu_q)\bigg]\;,
\lb{e}
\eq
and include the number density of the quark system
\bqn
n^{SLA}_q&=&
\frac{2N_c}{(2\pi)^3}\int d^3k 
\bigg[f^{SLA}_q(T,J^E_1,\mu_q)-{\bar f}^{SLA}_q(T,J^E_1,\mu_q)\bigg]\nonumber\\
&-&T\frac{\partial J^E_1}{\partial\mu_q}\frac{2N_c}{(2\pi)^3}\int d^3k 
\bigg[f^{SLA}_q(T,J^E_1,\mu_q)+{\bar f}^{SLA}_q(T,J^E_1,\mu_q)\bigg]\;,
\lb{dens}
\eqn
where
\bqn
f^{SLA}_q(T,\mu_q,J^E_1)=\frac{1}{\ex^{\beta(E+TJ^E_1-\mu_q)}+1}\;\;&{\rm and}&\;\;
{\bar f}^{SLA}_q(T,\mu_q,J^E_1)=\frac{1}{\ex^{\beta(E+TJ^E_1+\mu_q)}+1}\;
\lb{nJE}
\eqn
($q=u,d,s$), $E=\sqrt{k^2+m^2_q}$, $\beta=1/T$, and $J^E_1\equiv V_1/2T$ is the exponent of the 
Polyakov loop in the fundamental representation, where  
\bq
V_1=\int^\beta_0d\tau(1-\tau T)\int^\infty_0\xi d\xi D^E_1(\sqrt{\xi^2+\tau^2})\;.
\lb{V1}
\eq
is the large distance static ${\rm q\bar q}$ potential \cite{Si5,Si6,DiG}.

 The pressure and energy density of gluons are given by
\bq
p^{SLA}_{gl}=\frac{(N_c^2-1)}{3}\frac{2}{(2\pi)^3}\int d^3k
\frac{k}{\ex^{\beta(k+T{\tilde J}^E_1)}-1}
\lb{pgl}
\eq
and
\bq
\varepsilon^{SLA}_{gl}=3\;p_{gl}-T^2\frac{\partial\tilde{J}^E_1}{\partial T}(N_c^2-1)
\frac{2}{(2\pi)^3}\int d^3k\frac{1}{\ex^{\beta(k+T{\tilde J}^E_1)}-1}\;,
\lb{egl}
\eq
where ${\tilde J}^E_1=\frac{9}{4}J^E_1$ is the exponent of the Polyakov loop in the adjoint 
representation. 
 In Eqs.(\ref{nJE}), (\ref{pgl}) and (\ref{egl}), when $V_1=0$ we recover the ordinary Fermi 
and Bose gases. 
 In order to give Eqs.(\ref{p})-(\ref{egl}) in its most general forms, it was assumed 
that $V_1$ is, in principle, a function of temperature and chemical potential. 
 However, according to the parametrization given by Eq. (\ref{V1}),  $V_1$ does not depend 
on the chemical potential.  
 As pointed out in \cite{ST1}, the expected $\mu$-dependence of $V_1$ should be weak 
for values of $\mu$ much smaller than the scale of vacuum fields (which is of the order 
of $\sim 1.5$ GeV) and is partially supported by the lattice simulations \cite{Dor}.
 As in \cite{Bal,Bur,Fla}, we take $V_1$ independent of the chemical potential, so
$\;\partial J^E_1/\partial\mu_q=0\;$ in Eqs. (\ref{e}) and (\ref{dens}).

 In order to take into account the presence of electrons to keep the quark matter in 
$\beta$-equilibrium and with charge neutrality, we also include the equations for the 
pressure, energy density and number density of electrons given by
\bq
p_{\rm e}=\frac{1}{3}\frac{2}{(2\pi)^3}\int d^3k\frac{k^2}{E_\ex}
[f_{\rm e}(T,\mu_{\rm e})+{\bar f}_{\rm e}(T,\mu_{\rm e})]\;,
\lb{pel}
\eq
\bq
\varepsilon_{\rm e}=\frac{2}{(2\pi)^3}\int d^3k\;E_\ex
[f_{\rm e}(T,\mu_{\rm e})+{\bar f}_{\rm e}(T,\mu_{\rm e})]\;,
\lb{eel}
\eq
\bq
n_{\rm e}=\frac{2}{(2\pi)^3}\int d^3k
[f_{\rm e}(T,\mu_{\rm e})-{\bar f}_{\rm e}(T,\mu_{\rm e})]\;,
\lb{nel}
\eq
where
\bqn
f_{\rm e}(T,\mu_{\rm e})=\frac{1}{\ex^{\beta(E_\ex-\mu_{\rm e})}+1}\;\;&,&\;\;
{\bar f}_{\rm e}(T,\mu_{\rm e})=\frac{1}{\ex^{\beta(E_\ex+\mu_\ex)}+1}\;
\lb{fdel}
\eqn
and $E_\ex=\sqrt{k^2_\ex+m^2_\ex}$ .

 The composition of SQM is maintained in $\beta$-equilibrium with respect to weak 
interactions and in electric charge neutrality. 
 The weak interactions reactions are given by
\bq
d\rightarrow u+\ex+{\bar\nu}_\ex
\lb{due}
\eq
and
\bq
s\rightarrow u+\ex+{\bar\nu}_\ex\;.
\lb{sue}
\eq
 As pointed out in \cite{Far}, the neutrino gas is so dilute that it play no role in the 
dynamics of the system.
 So, by neglecting the neutrino chemical potential, the chemical equilibrium equations are 
given by 
\bq
\mu_d=\mu_u+\mu_\ex\;
\lb{mud}
\eq
and
\bq
\mu_s=\mu_d\;.
\lb{mus}
\eq
 The overall charge neutrality requires that
\bq
\frac{1}{3}(2n^{SLA}_u-n^{SLA}_d-n^{SLA}_s)-n_\ex=0\;.
\lb{chn}
\eq
 By numerically solving Eqs. (\ref{mud})-(\ref{chn}), for each value of the input 
total density 
\bq
n=n^{SLA}_u+n^{SLA}_d+n^{SLA}_s+n_\ex\;,
\lb{nT}
\eq
the unknown chemical potentials $\mu_{\rm u}$, $\mu_{\rm d}$, $\mu_{\rm s}$ and 
$\mu_{\rm e}$ are determined for fixed values of $T$, $G_2$ and $V_1$. 
 However, our calculation here is slightly different from that in \cite{Fla}, as 
explained below.

 The total pressure and energy density of the quark-gluon system, including electrons 
are given by
\bq
p=p^{SLA}_{gl}+\sum_{q=u,d,s}p^{SLA}_{q}-\Delta|\varepsilon_{vac}|+p_{\rm e}\;,
\lb{pqgl}
\eq
\bq
\varepsilon=\varepsilon^{SLA}_{gl}+\sum_{q=u,d,s}\varepsilon^{SLA}_{q}+
\Delta|\varepsilon_{vac}|+\varepsilon_{\rm e}\;,
\lb{eqgl}
\eq
where 
\bq
\Delta|\varepsilon_{vac}|=\frac{11-\frac{2}{3}N_f}{32}\Delta G_2\;,
\lb{dvac}
\eq
is the vacuum energy density defference between confined and deconfined phases in terms of 
the respective difference between the values of the gluon condensate, 
$\Delta G_2=G_2(T<T_c)-G_2(T>T_c)\simeq \frac{1}{2}G_2$ \cite{ST1,ST2}, and $N_f$ is the 
number of flavors.

 We follow the same line of \cite{Far} to investigate the behavior of the SQM at zero 
temperature\footnote{In reality, we take $T=0.001$ GeV in Eqs. (\ref{p}-\ref{nel}) which 
is a good approximation or, alternatively, by using Eqs.(\ref{pqT0}) - (\ref{nqT0}).} 
and total pressure by solving the above equations for constant values of the energy per 
baryon, $E/A=\varepsilon/n_A$, where $n_A=(n_u+n_d+n_s)/3$ is the baryon number density.
 We perform our calculation in the $m_u=m_d=0$ approximation\footnote{For our purposes 
here, it is irrelevant if electrons are assumed massless or not.} (in Sec. \ref{res}, 
larger values of $m_u$ and $m_d$ are speculated in order to look for strangeness 
excess and negative electric charge possibilities). 
 By using the constant $E/A$ constraint in the above equations, $m_s$ and 
$\Delta|\varepsilon_{vac}|$ (or $G_2$) are determined for fixed values of $V_1$ in the 
range $0\leq V_1\leq0.5$ GeV (where $V_1=0.5$ GeV is the value of $V_1$ at $T=T_c$ 
obtained from lattice investigations \cite{KaZ}).
 As in \cite{Bur,Bal,Fla}, our calculation here is made for $V_1$ constant, so  
$\;T^2\partial J^E_1/\partial T=-V_1/2\;$ in Eq. (\ref{e}). 

\subsection{\bf Quark matter at ${\bf T=0}$ and constant ${\bf V_1}$.}
\lb{eosT0}

 For pedagogical purposes, we show the previous equations for the quark system at zero 
temperature and constant $V_1$ for the general case of nonzero quark masses. 
 Zero temperature implies that 
\bq
f_q^{SLA}(T,\mu_q,J^E_1)\T0lim\Theta(\mu_q-E-TJ^E_1)
\lb{T0lim}
\eq
and Eqs.(\ref{p})-(\ref{dens}) lead to
\bq
p_q^{SLA}=\frac{N_c}{3\pi^2}\Bigg\{\frac{k_q^3}{4}\sqrt{k_q^2+m_q^2}-
\frac{3}{8}\;m_q^2\bigg[k_q\sqrt{k_q^2+m_q^2}-m_q^2\;\ln\bigg(\frac{k_q+\sqrt{k_q^2+m_q^2}}{m_q}\bigg)
\bigg]\Bigg\}\;,
\lb{pqT0}
\eq

\bqn
\varepsilon_q^{SLA}&=&\frac{N_c}{\pi^2}\Bigg\{\frac{k_q^3}{4}\sqrt{k_q^2+m_q^2}+
\frac{m_q^2}{8}\;\bigg[k_q\sqrt{k_q^2+m_q^2}-m_q^2\;\ln\bigg(\frac{k_q+\sqrt{k_q^2+m_q^2}}{m_q}
\bigg)\bigg]\nonumber\\
&+&\frac{V_1}{2}\;\frac{k_q^3}{3}
\Bigg\}
\lb{eqT0}
\eqn
and 
\bq
n_q^{SLA}=\frac{N_c}{\pi^2}\;\frac{k_q^3}{3}\;,
\lb{nqT0}
\eq
where
\bq
k_q=\sqrt{(\mu_q-V_1/2)^2-m_q^2}\;,\;\;\;\;(q=\rm{u,d,s}).
\lb{kq}
\eq 
 When $V_1=0$, the ordinary Fermi momentum $k_F$ is recovered.

\subsubsection{ Zero mass approximation}
\lb{eosT0m0}

\begin{figure*}[th]
\centerline{
\psfig{figure=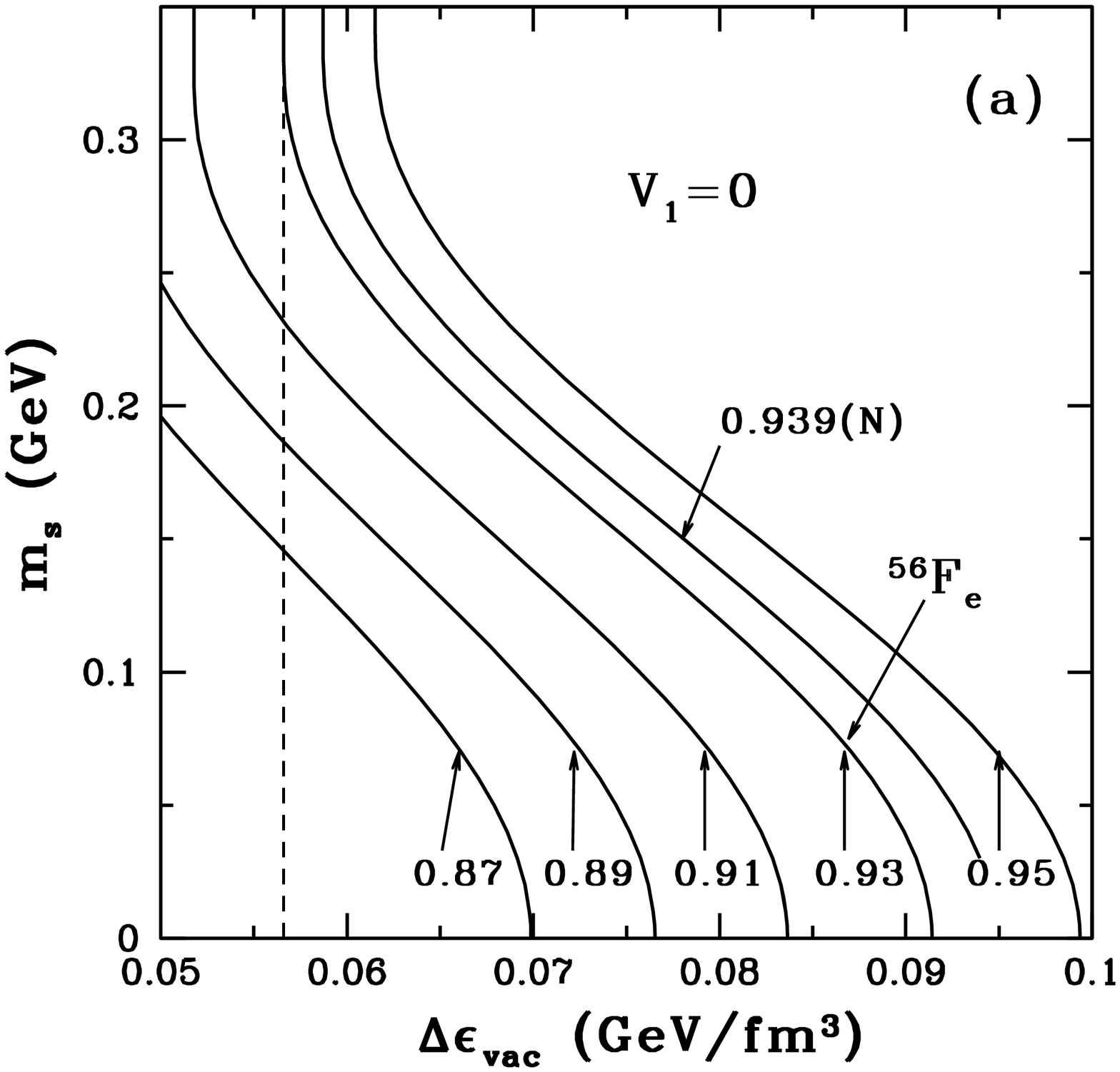,width=3.2truein,height=3.2truein}
\psfig{figure=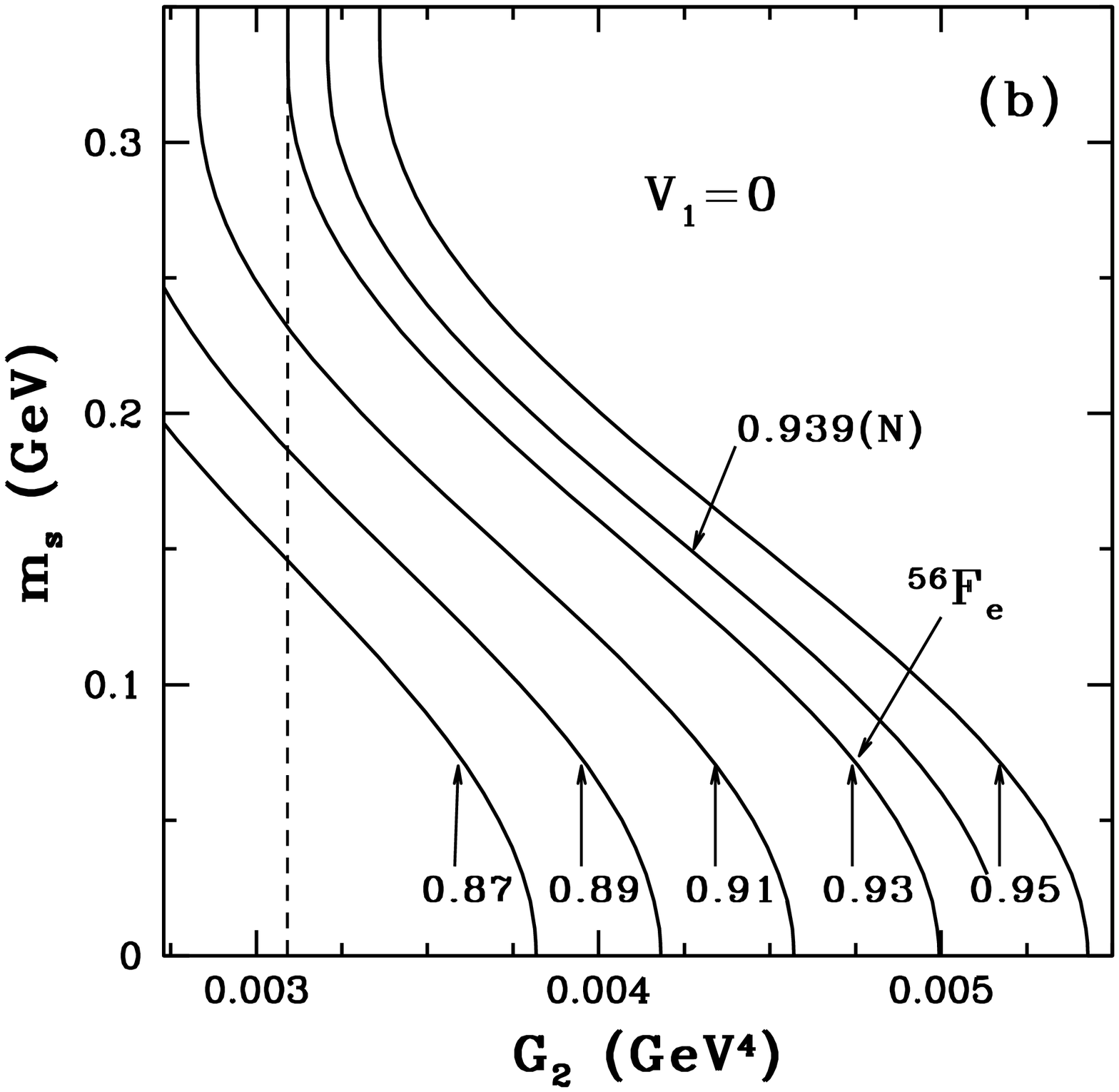,width=3.2truein,height=3.2truein}
\hskip .5in}
\centerline{
\psfig{figure=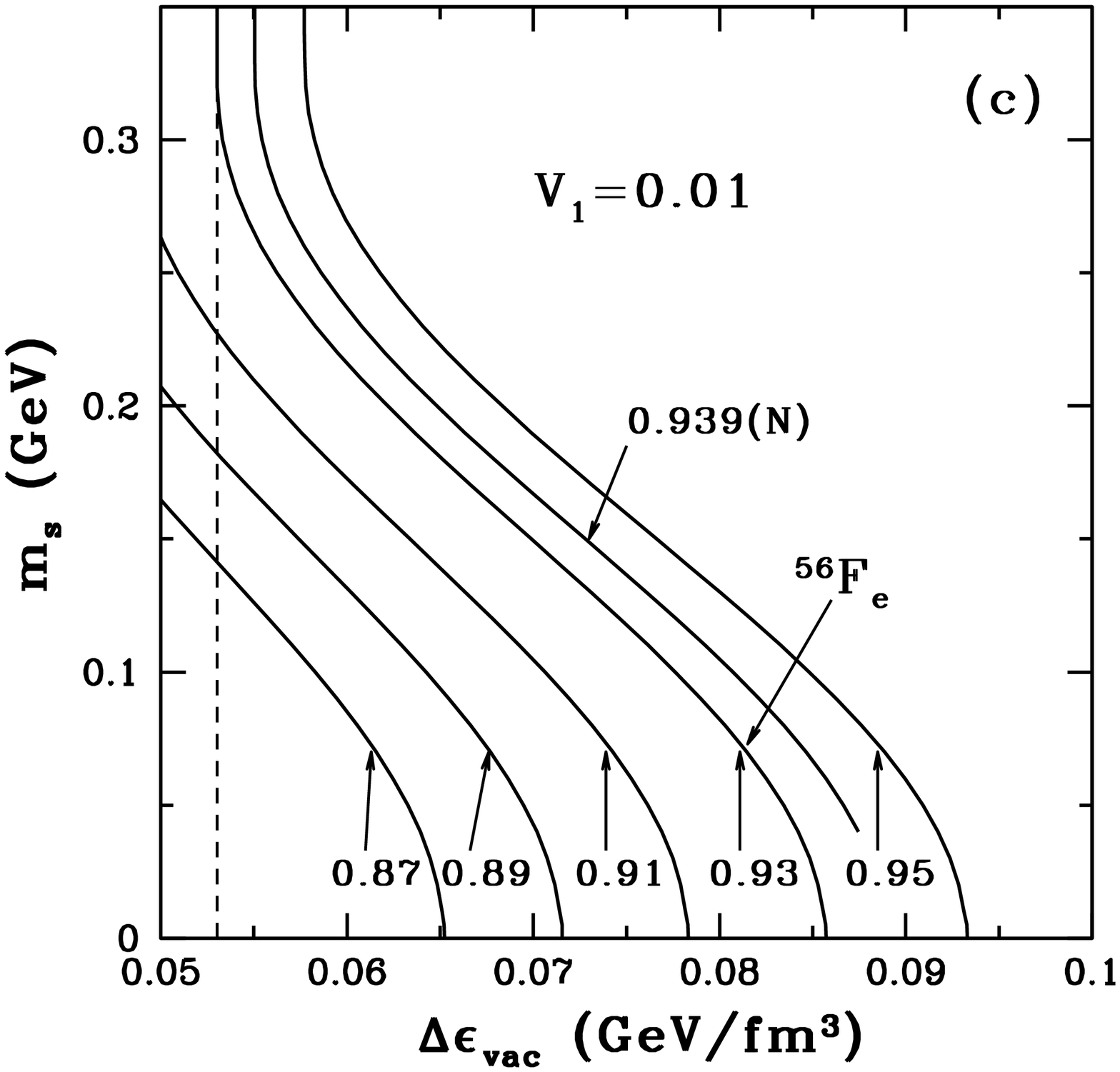,width=3.2truein,height=3.2truein}
\psfig{figure=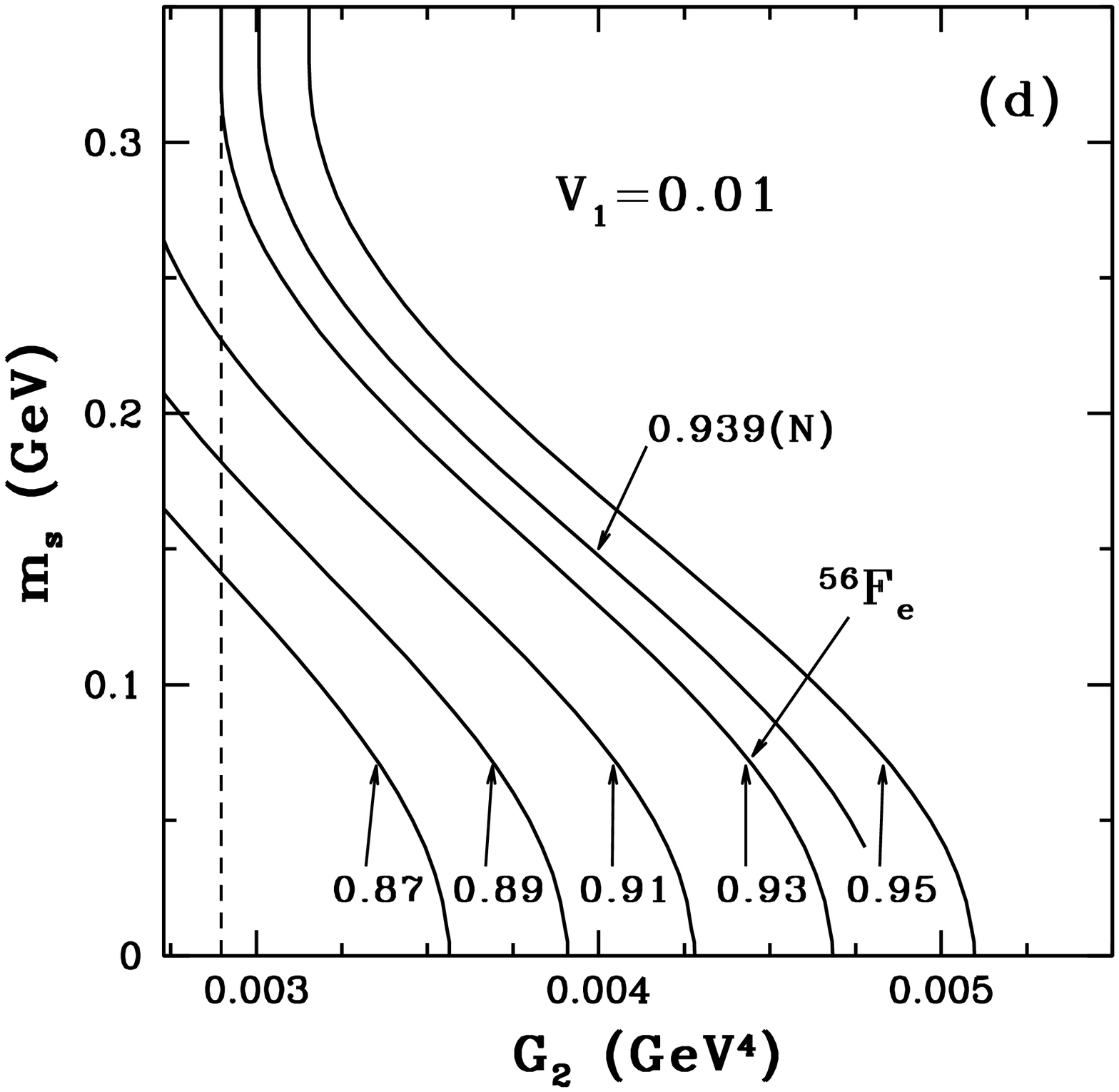,width=3.2truein,height=3.2truein}
\hskip .5in}
\caption{ 
}
\label{msevG2}
\end{figure*}

 In order to better understand the role of $G_2$ and $V_1$ in the study of stability of 
quark matter, it is instructive to apply the above equations to quark matter at zero 
temperature and pressure, assuming that all quark species are massless particles. 
 This simple case serve to help us to understand the behavior of the constant $E/A$ 
curves in Fig. \ref{msevG2}, as well as the shrinking of the stability window in panel (a) 
of Fig. \ref{fewin}, at $m_s=0$.
 For massless quarks, Eqs.(\ref{pqT0})-(\ref{nqT0}) (with $N_c=3$), by using Eq.(\ref{kq}), are reduced to
\bq
p_q^{SLA}=\frac{1}{4\pi^2}\;\bigg(\mu_q-\frac{V_1}{2}\bigg)^4\;,
\lb{pqT0m0}
\eq
\bqn
\varepsilon_q^{SLA}&=&\frac{1}{\pi^2}\bigg\{\frac{3}{4}\bigg(\mu_q-\frac{V_1}{2}\bigg)^4+
\;\frac{V_1}{2}\;\bigg(\mu_q-\frac{V_1}{2}\bigg)^3
\bigg\}
\lb{eqT0m0}
\eqn
and
\bq
n_q^{SLA}=\frac{1}{\pi^2}\;\bigg(\mu_q-\frac{V_1}{2}\bigg)^3.
\lb{nqT0m0}
\eq
 At total zero pressure, the sum of the quark pressures is balanced by the vacuum 
energy density,
\bq
\sum_q\frac{1}{4\pi^2}\;\bigg(\mu_q-\frac{V_1}{2}\bigg)^4-\Delta|\varepsilon_{vac}|=0\;,
\lb{}
\eq
and the energy density is
\bqn
\varepsilon&=&\sum_q\varepsilon_q+\Delta|\varepsilon_{vac}|\nonumber\\
&=&3\sum_q p_q+\frac{V_1}{2\pi^2}\sum_q\bigg(\mu_q-\frac{V_1}{2}\bigg)^3+\Delta|\varepsilon_{vac}|\nonumber\\
&=&\frac{3}{2}\;V_1\;n_A+4\;\Delta|\varepsilon_{vac}|
\lb{}
\eqn 
where (here, in this section) the baryon number density is $n_A=(n_u+n_d+n_s)/3$ for SQM and 
$n_A=(n_u+n_d)/3$ for nonstrange quark matter.
 Notice the presence of the additional term $(3/2)V_1n_A$ with respect to the corresponding 
expressions in the MIT Bag Model \cite{Mad}.
 Also noticed is that the sum of the quark pressures and energy densities are given in terms of 
$\Delta|\varepsilon_{vac}|$ (or $G_2$) and $V_1\;$ (differently from the MIT Bag Model where they 
are given solely in terms of the bag constant $B$). 
Now, let us particularize the above equations for two and three flavor quark matter.

\centerline{Two flavor}

 For a gas of $u$ and $d$ quarks, charge neutrality (neglecting the not important cotribution 
of electrons as in \cite{Mad}) requires that $n_d^{SLA}=2\;n_u^{SLA}$, from which it follows that 
$(\mu_d-V_1/2)=2^{1/3}(\mu_u-V_1/2)\;$.
 The two-flavor vacuum energy density (for $N_f=2$ in Eq.(\ref{dvac})) is 
$\Delta|\varepsilon_{vac}|_{ud}=(29/192)\;G_2$.
 Thus, the energy per baryon of the $ud$ system is 
\bqn
\bigg(\frac{E}{A}\bigg)_{\;ud}&=&
(1+2^{4/3})^{3/4}(4\pi^2)^{1/4}(\;\Delta|\varepsilon_{vac}|_{ud}\;)^{1/4}+\frac{3}{2}V_1\nonumber\\
&=&6.441\;(\;\Delta|\varepsilon_{vac}|_{ud}\;)^{1/4}+\frac{3}{2}V_1\nonumber\\
&=&4.016\;G_2^{\;1/4}+\frac{3}{2}V_1\;.
\lb{EAud}
\eqn

\centerline{Three flavor}

 The three flavor quark system (SQM) is naturally charge neutral, with 
$n_u^{SLA}=n_d^{SLA}=n_s^{SLA}\;$, $\mu_u=\mu_d=\mu_s=\mu\;$, and $n_\ex=\mu_\ex=0\;$. 
 The vacuum energy density (for $N_f=3$) is $\Delta|\varepsilon_{vac}|=(9/64)\;G_2\;$, 
and the energy per baryon becomes (using the previous notation for the SQM system, 
without subscripts) 
\bqn
\bigg(\frac{E}{A}\bigg)&=&
3^{3/4}(4\pi^2)^{1/4}(\;\Delta|\varepsilon_{vac}|\;)^{1/4}+\frac{3}{2}V_1\nonumber\\
&=&5.714\;(\;\Delta|\varepsilon_{vac}|\;)^{1/4}+\frac{3}{2}V_1\nonumber\\
&=&3.499\;G_2^{\;1/4}+\frac{3}{2}V_1\;.
\lb{EASQM}
\eqn
 In Eqs.(\ref{EAud}) and (\ref{EASQM}), for fixed $E/A$, the increase of $V_1(G_2)$ is 
compensated by the corresponding decrease of  $G_2(V_1)$. 
 As shown below in Fig. \ref{msevG2}, (at $m_s=0$ and a given $E/A$) the maximum value of $G_2$ is obtained 
for $V_1=0$\;.

 The energy per baryon of $^{56}F_\ex$ is 930.4 MeV, so in this simple analysis the 
stability of SQM relative to iron corresponds to $G_2<(0.266-0.428V_1)^4$. 
 As a result, we obtain $G_2<0.005\;{\rm GeV}^4$ for $V_1=0$, $G_2<0.002\;{\rm GeV}^4$ for 
$V_1=0.1\;$GeV, and so on, until a very small value of $G_2$ ($\gappl10^{-5}\;{\rm GeV}^4$) 
for $V_1=0.5\;$GeV. 
  This behavior explains the shrinking of the stability window at $m_s=0\;$ shown in panel (a) 
of Fig.\ref{fewin}.

Finally, from Eqs. (\ref{EAud}) and (\ref{EASQM}) we obtain
\bqn
\frac{(E/A)}{(E/A)_{\;ud}\;\;\;}&=&
\frac{3^{3/4}(4\pi^2)^{1/4}(\;\Delta|\varepsilon_{vac}|\;)^{1/4}\;+\;1.5\;V_1}
{(1+2^{4/3})^{3/4}(4\pi^2)^{1/4}(\;\Delta|\varepsilon_{vac}|_{ud}\;)^{1/4}\;+\;1.5\;V_1}\nonumber\\
&=&\frac{5.714\;(\;\Delta|\varepsilon_{vac}|\;)^{1/4}+\;1.5\;V_1}
{6.441\;(\;\Delta|\varepsilon_{vac}|_{ud}\;)^{1/4}+\;1.5\;V_1} 
\nonumber\\
&=&\frac{3.499\;G_2^{\;1/4}+\;1.5\;V_1}{4.016\;G_2^{\;1/4}+\;1.5\;V_1}\;.
\lb{SQMud}
\eqn
 It is evident that $(E/A)<(E/A)_{ud}$ (for the same values of $G_2$ and $V_1$ in SQM and ud systems). 
 From the last line of Eq.(\ref{SQMud}) it follows that $(E/A)/(E/A)_{ud}=0.87$ for $V_1=0$. 
 On the other hand, for the MIT Bag Model, with the correspondences   
$\Delta|\varepsilon_{vac}|=\Delta|\varepsilon_{vac}|_{ud}\equiv B$ and $V_1=0$, we obtain 
$(E/A)/(E/A)_{ud}=0.89$ as in \cite{Mad}. 

\section{Results}
\lb{res}

 We are concerned with the bulk properties of SQM and concentrate ourselves to investigate 
the stability with respect to the $^{56}F_\ex$ nucleus.
 In our investigation, $m_s$ enters as input parameter and 
$\Delta|\varepsilon_{vac}|$(or $G_2$) is determined for fixed values of $E/A$ and $V_1$. 
  By this way, as in \cite{Fla}, but with a different logic, we obtain a scenario for the model parameters, 
independently of what the results of lattice calculation may be. 
  We discuss the relations between the parameter values required for the SQM  stability and the 
values obtained by comparison with lattice predictions in \cite{ST1,ST2}.

\begin{figure*}[th]
\centerline{
\psfig{figure=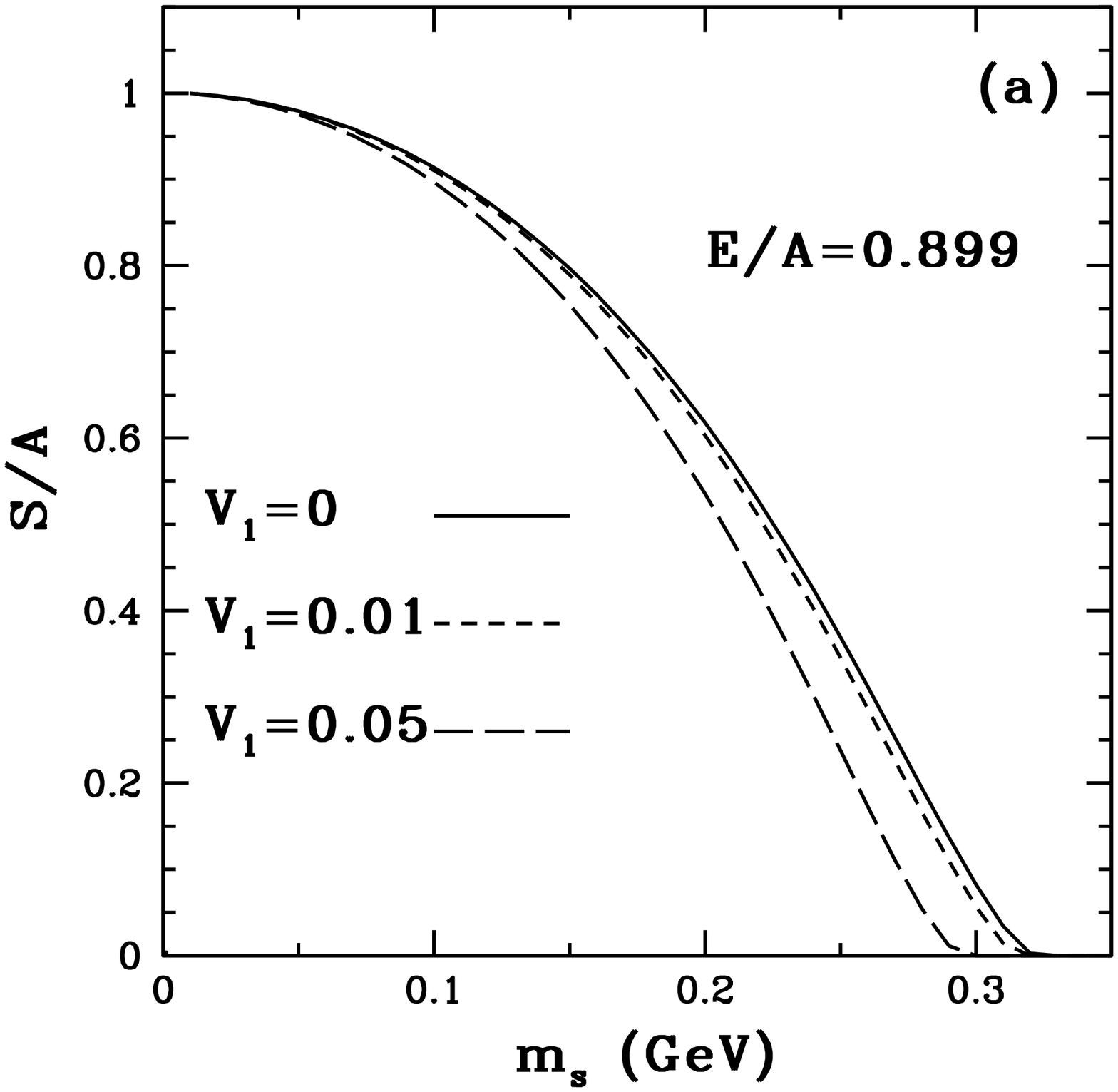,width=3.2truein,height=3.2truein}
 \psfig{figure=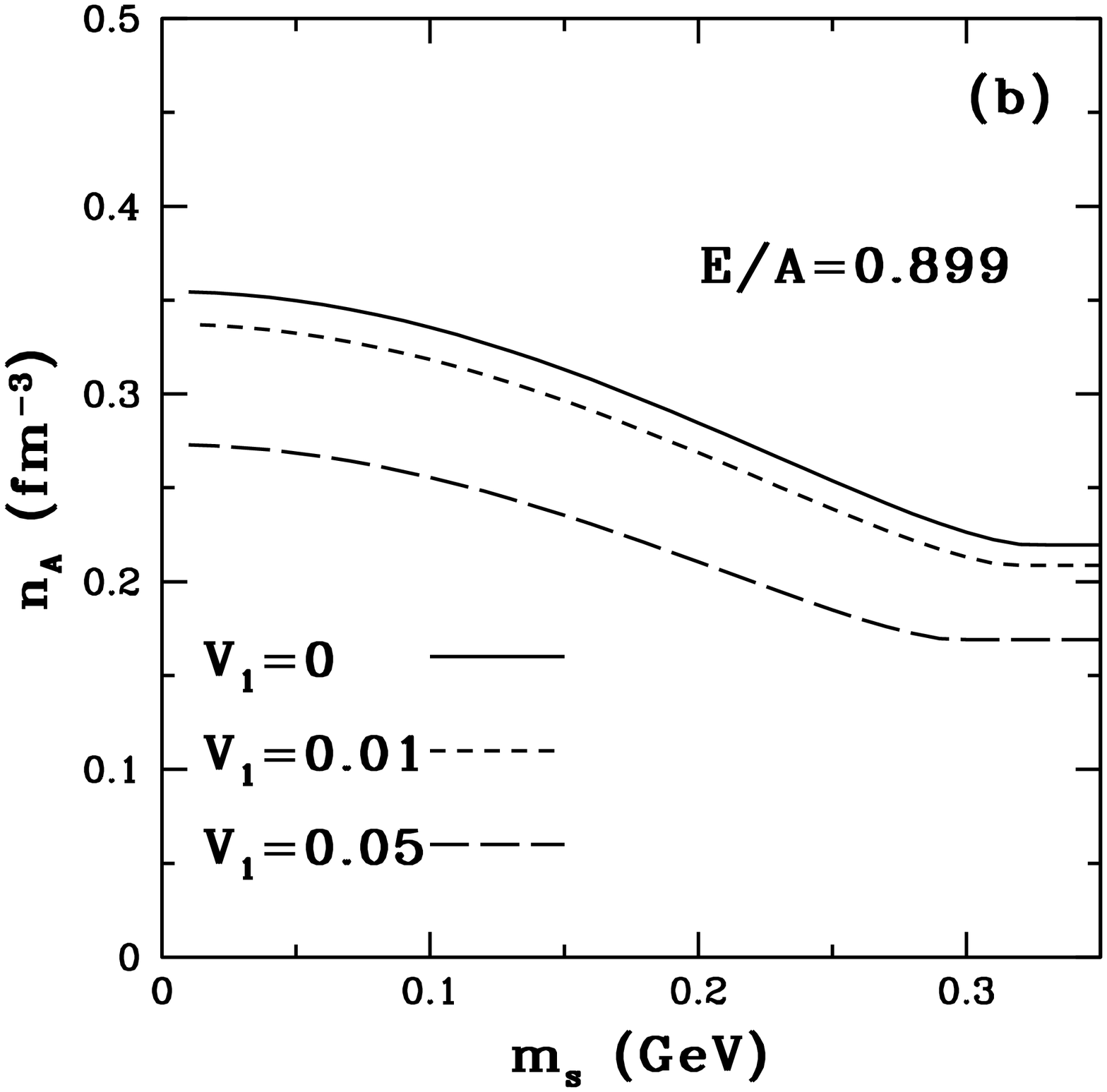,width=3.2truein,height=3.2truein}
\hskip .5in}
\centerline{ 
\psfig{figure=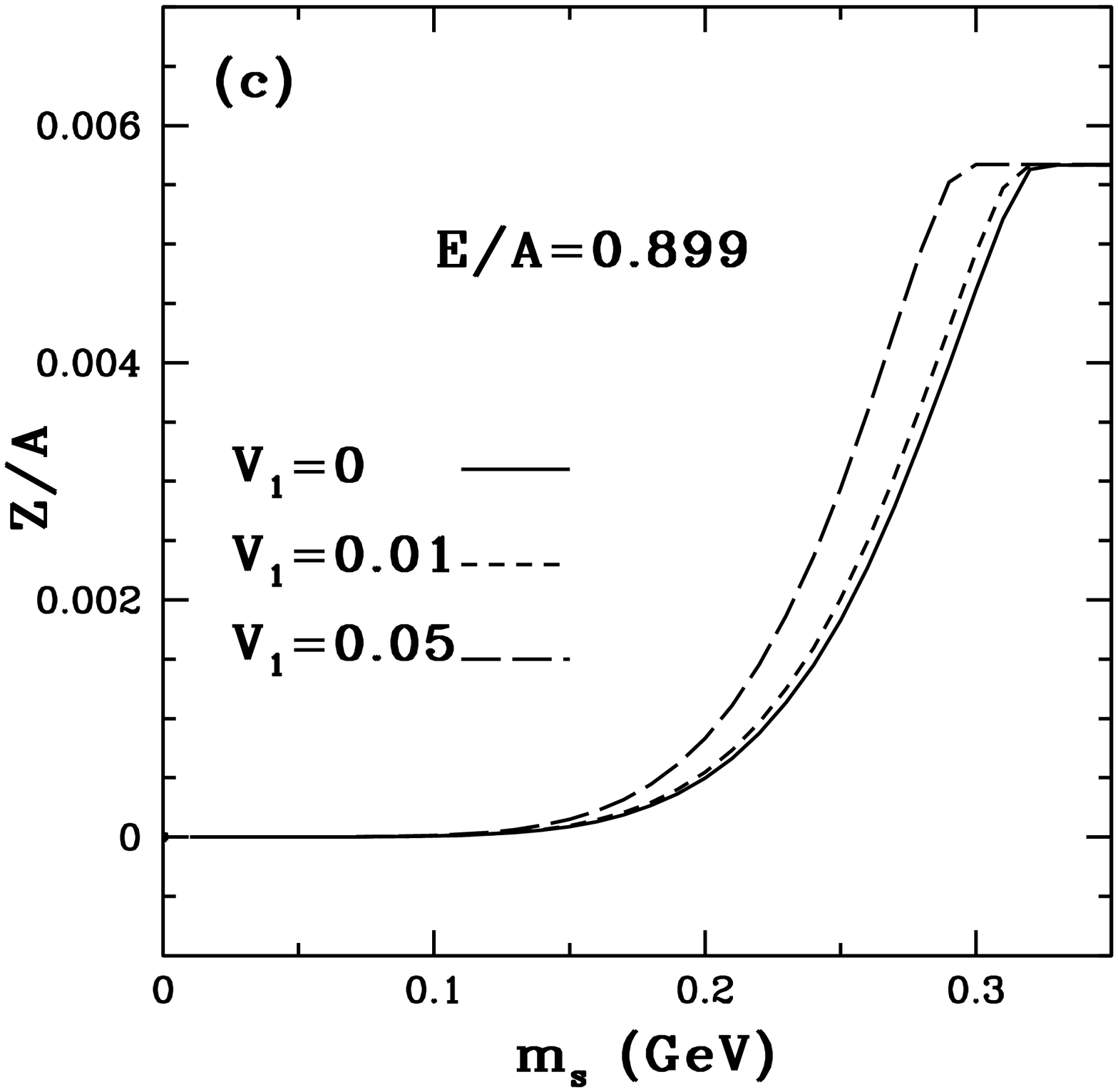,width=3.2truein,height=3.2truein}
\hskip .5in}
\caption{ 
}
\label{str}
\end{figure*}

 In Fig. \ref{msevG2}, the constant $E/A$ contours give $m_s$ vs $\Delta|\varepsilon_{vac}|$ 
(for the purpose of comparison with MIT Bag Model results in \cite{Far,KWW,Mad}) for 
different values of $V_1$. 
 In order to understand the role the gluon condensate, we use the relation between 
$\Delta|\varepsilon_{vac}|$ and $G_2$, given by Eq. (\ref{dvac}), to plot $m_s$ vs $G_2$ for 
the same values of $V_1$.
 The contours are very sensitive to the values of $V_1$, being shifted towards lower values of 
$\Delta|\varepsilon_{\rm vac}|$ and/or $G_2$ as shown for $V_1=0$ (panels (a) and (b)) and  
$V_1=0.01$ GeV (panels (c) and (d)).
 To the right of the $E/A=0.93$ GeV contour (in reality, 930.4 MeV corresponding to the energy per 
nucleon of $\;^{56}F_\ex$), SQM is unstable with respect to the iron nuclei. 
 The vertical line at the left of each panel is the limit of $\Delta|\varepsilon_{\rm vac}|$ 
and/or $G_2$ when $m_s$ becomes large, so the strangeness per baryon goes to zero (see panel (a)  
in Fig. \ref{str}).
 In this case, there is no distinction between strange and non-strange quark matter.
 Contours with $E/A<930.4$ MeV terminate at the crossing with the vertical line of 
$^{56}F_\ex$.
 For $V_1=0$, the results shown in panel (a) are numerically equivalent to the ones found in the case 
of the MIT Bag Model. 
 However, we remark that in the FCM the vacuum energy difference $\Delta|\varepsilon_{\rm vac}|$ is 
essentially a nonperturbative quantity given in terms of the gluon condensate.
 Also shown is the $E/A=0.939$ GeV contour corresponding to the nucleon mass. 
  
 Stability window of the SQM is the region of allowed values of $m_s$ and 
$\Delta|\varepsilon_{\rm vac}|$ (or $G_2$) where the energy per particle is lower than the 
one of $^{56}F_\ex$ (bounded by the $E/A=0.93$ GeV contour and the respective vertical line). 
 In the FCM, the stability of SQM depends on the values of $V_1$ and/or $G_2$. 
 For a given value of $E/A$, the higher $V_1$, the lower $G_2$ (cf. Eq.(\ref{EASQM}) for 
the case $m_s=0$).
 Moreover, even for $V_1=0$ (for which the contours present the maximum $G_2$ at $m_s=0$), 
the values of $G_2$ within the stability window are lower than $0.006-0.007\;{\rm GeV}^4$ 
obtained from lattice data on the critical temperature \cite{ST1,ST2}.
 The possibility of the SQM be more bound than $^{56}F_\ex$ is realized only for $G_2<0.005\;{\rm GeV}^4\;$.
 This has also been the case for $m_u=5$ MeV, $m_d=7$ MeV and $m_s=150$ MeV for which we have shown that 
$G_2<0.0041\;{\rm GeV}^4$ for the existence of stable SQM in strange star surfaces \cite{Fla}. 
  Even if we take stability with respect to the nucleon mass ($E/A=0.939\;$GeV), instead of 
$^{56}F_\ex$, the values of $G_2$ remain lower than the one in \cite{ST1,ST2}.

 Fig. \ref{str} shows for the given value of $E/A$ (the same as in \cite{Far}) the strangeness 
per baryon (defined as in \cite{Far}) in panel (a) as function of $m_s$ for some values $V_1$. 
 The strangeness per baryon is always lower then unity, going to zero at $m_s\sim0.3$ GeV (however, 
strangeness excess might be possible for nonzero $m_u$ and/or $m_d$ as shown in panel (a) of 
Fig. \ref{strqe}).
 For the same values of $V_1$, panel (b) shows the decrease of the baryon number density from 
its maximum at $m_s=0$ until a constant value around $m_s\sim0.3$ GeV.
 We must be aware that for larger values of  $V_1$ and at some value of $m_s$, the baryon number 
density $n_A$ might becomes lower than a critical value (if it exists) at which the phase transition 
takes place.
 However, the determination of such a critical value is not the scope of the present work.

 We have also calculated the hadronic electric charge per baryon, 
\bq
\frac{Z}{A}=\frac{2n_u-n_d-n_s}{n_u+n_d+n_s}=\pm\frac{n_\ex}{n_A}\;,
\lb{ZA}
\eq
 shown in panel (c).
 For $m_s=0$ the equilibrium configuration is given by an equal number of $u$, $d$ and $s$ 
quarks ($n_u=n_d=n_s$ and $n_{\rm e}=0$) with zero electric charge.
 When $m_s$ and/or $V_1$ grow, the system develops a positive hadronic electric charge. 
 For large $m_s$, the hadronic electric charge per baryon saturates at a constant value 
which also depends on $V_1$ (cf. panel (b) of Fig. \ref{festr}). 
 For $V_1=0$, this saturation point is $\sim0.0056$ at $m_s\sim0.3$ GeV as in \cite{Far}.

 Given that all $E/A<0.93$ GeV contours are within the stability window, it is instructive 
to consider some features of SQM at $E/A=0.93$ GeV of $^{56}F_\ex$. 
 In panel (a) of Fig. \ref{fewin} we show that the overall effect of the confining forces is 
to shift the stability windows towards lower values of $G_2$ for increasing values of $V_1$.
 The windows not only narrow as $m_s$ grows at a fixed $V_1$, but they also narrow as $V_1$ 
grows at the same value of $m_s$.
 In particular, for $V_1=0.5$ GeV the vertical line is located at a negligible 
value of $G_2$ and the stability window width is very small (not visible in the scale of the figure) .  
 From the locations (at fixed values of $V_1$) of the constant $E/A$ contours and the respective 
vertical lines on the horizontal axis at $m_s=0$, where each window presents its maximum width, 
we construct two plots $V_1$ vs $G_2$ as shown in panel (b). 
 The region between the dashed and solid curves illustrates the decrease of the stability window 
width with $V_1$.
 In panel (c), we show the baryon number density for $0\leq V_1\leq0.5\;$GeV. 
 For $V_1=0.5$ GeV, it is nearly zero.
 So, as we have observed above, due to the decreasing of $n_A$ it might happens that a 
quark-hadron phase transition occurs at some value of $V_1$ and $m_s$.

  Fig. \ref{festr} shows the strangeness per baryon (in panel (a)) and the hadronic electric charge 
per baryon (in panel (b)), at the energy per baryon of $\;^{56}F_\ex$, as function of $m_s$ for  
$V_1=0$ and $V_1=0.5$ GeV.
 For all values of $V_1$ in the region $0\leq V_1\leq0.5$ GeV and $m_s$ in the region 
$0\leq m_s\gappl0.35\;$GeV, the strangeness per baryon is always less than unity. 
 Depending on the values of $V_1$, the saturation of the hadronic electric charge per baryon is  
between $0.0056$ for $V_1=0$ and $\sim0.006$ for $V_1=0.5$ GeV (this variation is not visible in 
panel (c) of Fig. \ref{str} because of the low values of $V_1$).

 Let us now consider the question of the strangeness excess. 
 We have performed our calculation assuming that $m_u=m_d=0$.
 However, for nonzero $m_u$ and/or $m_d$, strangeness excess can be obtained  which is 
more sensitive to $m_d$ than it is to $m_u$. 
 For the usual values of $u$ and $d$ quark masses, $m_u=5$ MeV and $m_d=7$ MeV, the strangeness 
excess is less than 1\%, but it can be larger for larger $m_u$ and $m_d$. 
 As an ilustrative example, we have (speculatively) extrapolated $m_u$ and $m_d$ beyond its 
usual values in order to obtain a strangeness excess around 9-14 \% as shown in panel (a) of 
Fig. \ref{strqe},  for the given quark masses, $V_1$ and $E/A$. 
 Generally speaking, this excess only occurs for low values of $m_s$ ($\gappl0.02\;$GeV) and 
large values of $V_1$ which, in turn, correspond to very small values of 
$G_2$ ($\sim10^{-6}\;{\rm GeV}^4$) (the increase of $u$ and $d$ quark masses also shifts the 
$E/A$ contours towards lower values of $G_2$).

 Correspondingly, we also speculate the possibility of negative hadronic electric charge.
 The change of the hadronic electric charge is more sensitive to $m_u$ and $m_s$ than it is 
to $m_d$.
 For large values of $V_1$, it can happen that strange quarks are more abundant than the massless 
$u$ and $d$ quarks in the region of  small values of $m_s$.
 So, negative hadronic electric charge appears to be allowed for large values of $V_1$, $m_u$ 
and $m_d$ (as for the case of strangeness excess), but for small $m_s$, as shown in panel (b). 
 In this case, instead of electrons, a sea of positrons neutralizes the negative hadronic 
electric charge. 
 We also remark that, at the same values of $V_1$, $m_u$ and $m_d$, larger values of $S/A$ and 
lower (negative) $Z/A$ are allowed for $E/A<0.899$ GeV and $m_s$ in the region 
$0\leq m_s\gappl0.02\;$GeV.

 Summarizing the above results, strangeness excess and negative hadronic electric charge per 
baryon are realized only for large values of $m_u$ and/or $m_d$ and $V_1$ (say, $V_1\gappr0.3$ GeV), 
but for values of $G_2$ much lower than the one in \cite{ST1,ST2}. 
 For $m_u=m_d=0$ as well as for the usual $m_u=5$ MeV and $m_d=7$ MeV, we observe that the values 
of the model parameters obtained in the present paper do not favor the existence of neither 
nonnegative nor negative SQM with energy per baryon lower than the one of the $^{56}F_\ex$ nucleus.

\section{Final remarks and conclusions}
\lb{frem}

 In this work we have investigated the bulk properties of SQM by using the quark-gluon 
plasma EOS derived in the FCM nonperturbative approach \cite{Si6}. 
 The important parameters of the model are the gluon condensate $G_2$ (which enters the 
EOS through the vacuum energy difference $\Delta|\varepsilon_{\rm vac}|$ between confined 
and deconfined phases) and the large distance ${\rm q}\bar{\rm q}$ interaction potential $V_1$. 
 The results have shown that confinement plays an important role for the stability of SQM. 
 
 We have performed the calculation in the $m_u=m_d=0$ approximation and assumed SQM in 
$\beta$-equilibrium and charge neutrality, at zero temperature and pressure. 
 We have been mainly concerned with the absolute stability of SQM with respect to $^{56}F_\ex$ 
nucleus.
 In order to look for stability windows of SQM, we have started our investigation  by 
drawing contours of constant energy per baryon, $E/A$, in the 
$m_s$ vs $\Delta|\varepsilon_{\rm vac}|$ (and/or $m_s$ vs $G_2$) plane for fixed 
values of $V_1$. 
 Strangeness and hadronic electric charge have also been considered.
 Our study revealed remarkable features which we summarize as follows.

 The general trend is that the SQM stability is very sensitive to the values of the model 
parameters responsible by the confining forces.
 A remarkable aspect is that the behavior of the stability window strongly depends on the 
values of $V_1$. 
  For increasing values of this parameter, the constant $E/A$ contours and also the respective 
stability windows as a whole are shifted towards lower and lower values of 
$\Delta|\varepsilon_{\rm vac}|$ (and/or $G_2$) until to $\sim0$ at $V_1=0.5$ GeV.
  Moreover, the width of the stability windows diminish when $V_1$ becomes larger.
 At $m_s=0$, it has the maximum width between $G_2=0.003\;{\rm GeV}^4$ and $G_2=0.005\;{\rm GeV}^4$ 
for $V_1=0$ and a nearly zero width at $G_2\sim0$ for $V_1=0.5$ GeV. 
 This amounts to say that absolutely stable SQM would exists, in principle, for $0\leq V_1\leq0.5$ 
GeV (although somewhat problematic at $V_1=0.5$ GeV due to the smallness of the corresponding value 
of $G_2$).
 However, a striking point is that the values of $G_2$ are lower than the one in the range 
$0.006-0.007\;{\rm GeV}^4$ obtained from comparison with the lattice data at the critical temperature 
\cite{ST1,ST2}. 
 This point puts a severe restriction for the existence of absolutely stable SQM with respect to 
$^{56}F_\ex$.

 We have also calculated the strangeness per baryon and the hadronic electric charge per baryon. 
 As $m_s$ grows, the strangeness per baryon decreases from $S/A=1$ at $m_s=0$ to $S/A=0$ 
for some value of $m_s$ which depends on $E/A$ and $V_1$.
 Correspondingly, the hadronic electric charge per baryon is always nonnegative, rising from 
$Z/A=0$ up to a constant value which depends on the value of $V_1$. 
 Another remarkable feature (in the $m_u=m_d=0$ approximation) is that $S/A\leq1$ and $Z/A\geq0$ 
for all values of $m_s$ within the stability window. 
  
 In the attempt to find strangeness excess and negative hadronic electric charge, we have 
observed that $S/A>1$ and $Z/A<0$ appear to be allowed only for very large values of $m_u$ and/or 
$m_d$ (beyond the usual ones), but for small $m_s$ ($\gappl0.02$ GeV) and large $V_1$ (say, between 0.3 GeV and 
0.5 GeV).
 However, the corresponding values of $G_2$ remain lower than the one in \cite{ST1,ST2}, as in 
the case of  $m_u=m_d=0$.
 
 From the above, in the context of the FCM approach, it appears that the values of the model 
parameters obtained in our investigation do not favor the existence of absolutely stable SQM. 
 Of course the above results depend on the constant $V_1$ assumption, in the present 
work. 

\begin{figure*}[th]
\centerline{
\psfig{figure=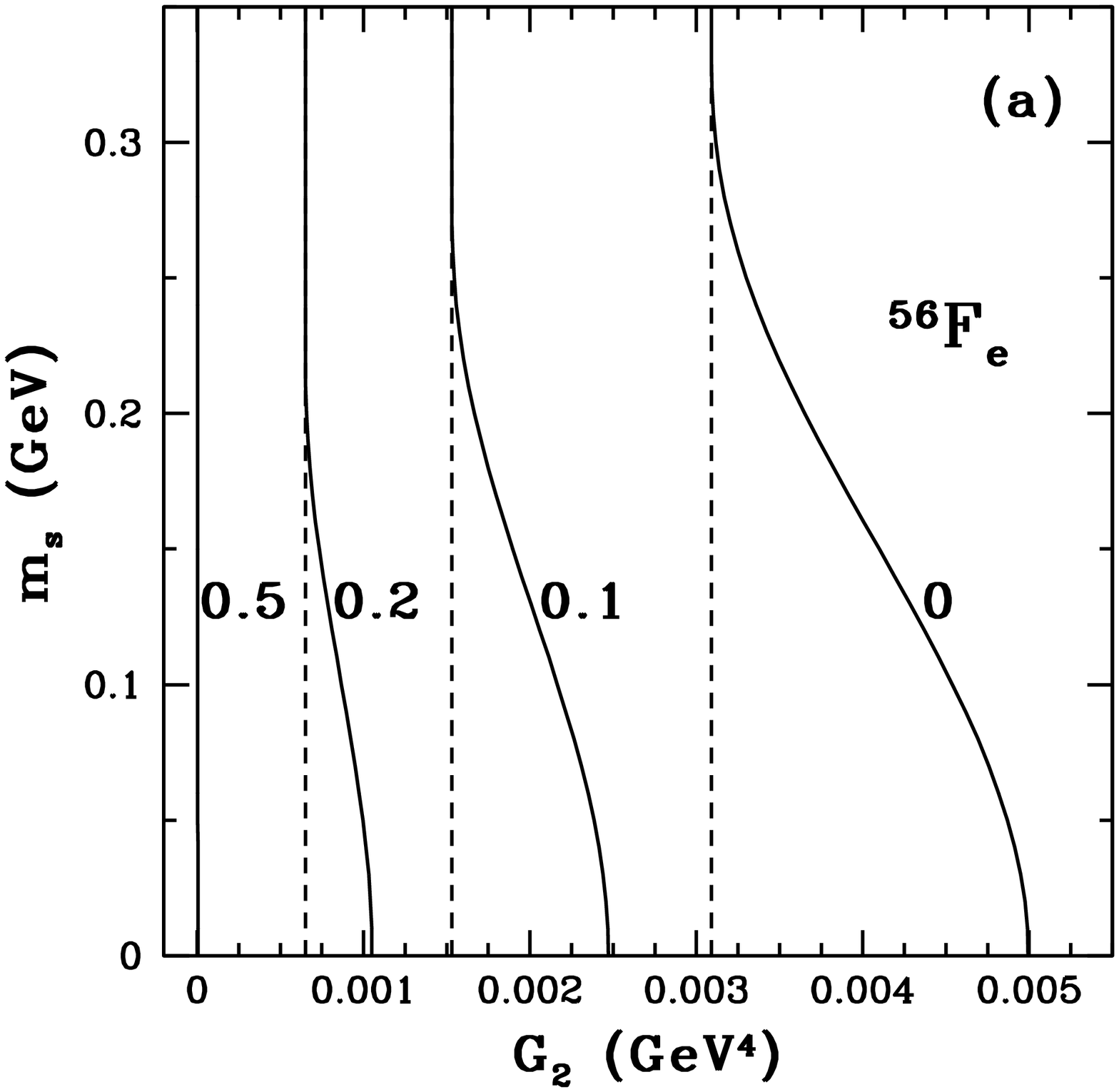,width=3.2truein,height=3.2truein}
   \psfig{figure=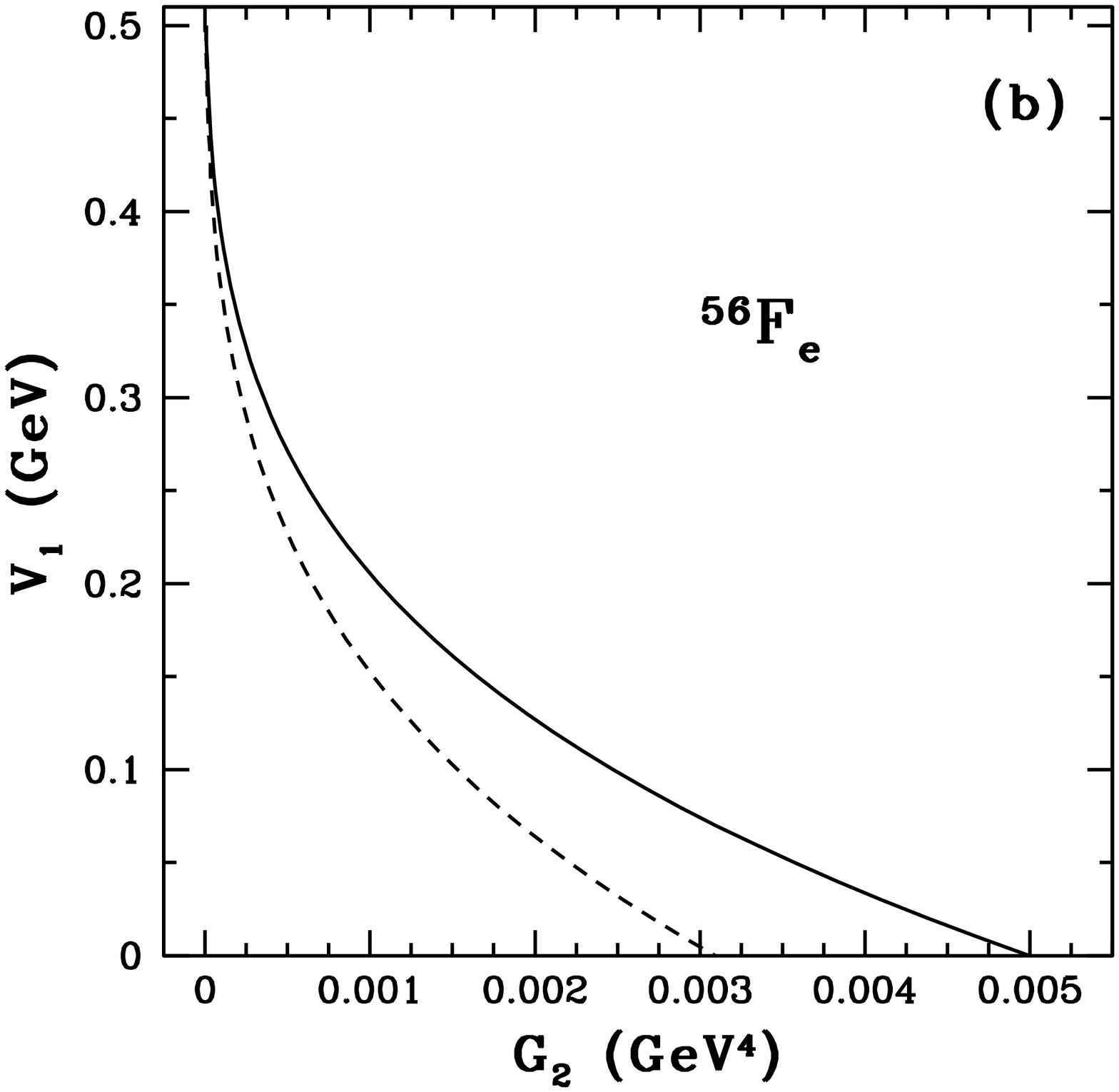,width=3.2truein,height=3.2truein}
\hskip .5in}
\centerline{
  \psfig{figure=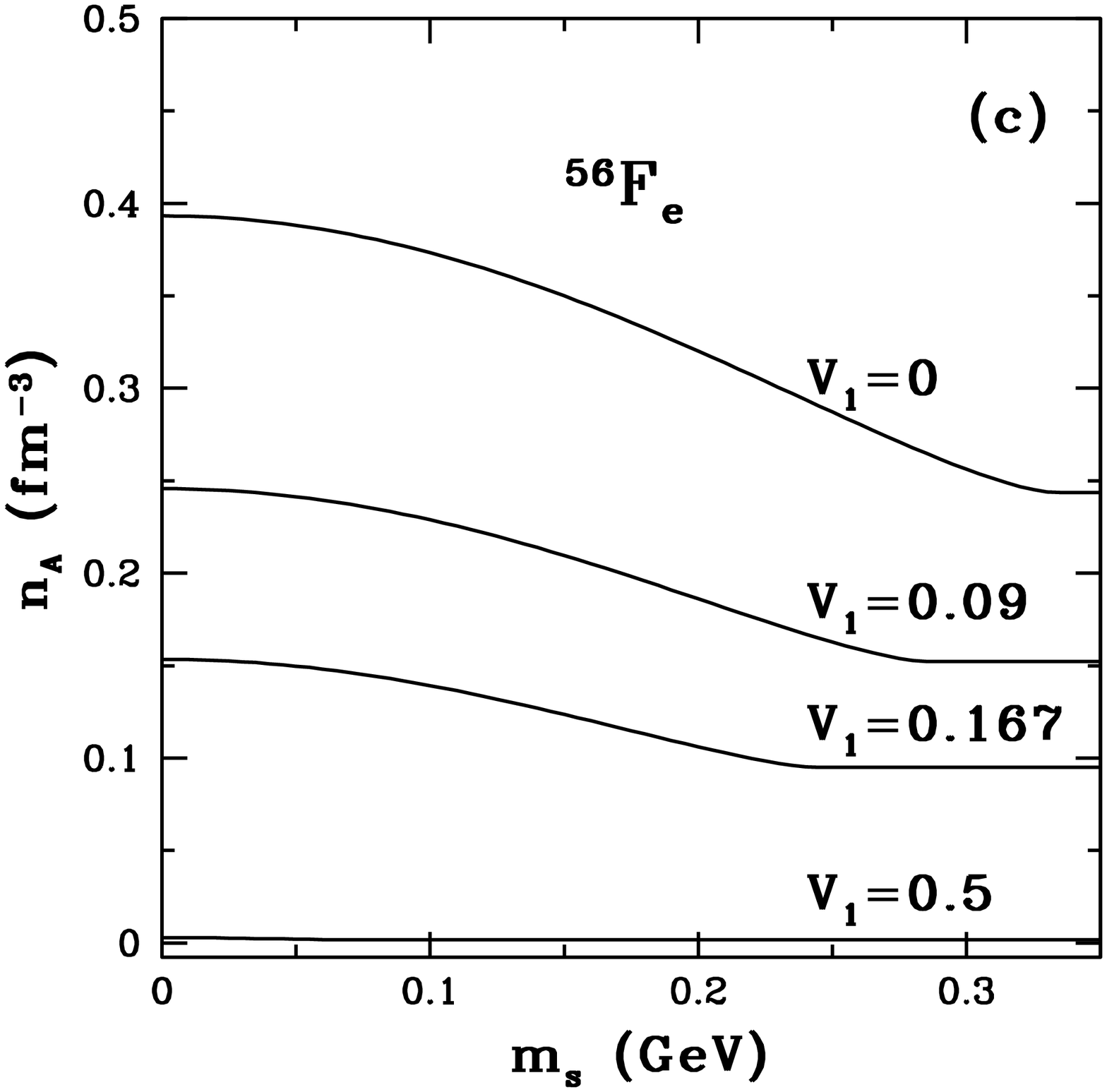,width=3.2truein,height=3.2truein}
\hskip .5in}
\caption{ 
}
\label{fewin}
\end{figure*}

\begin{figure*}[th]
\centerline{
\psfig{figure=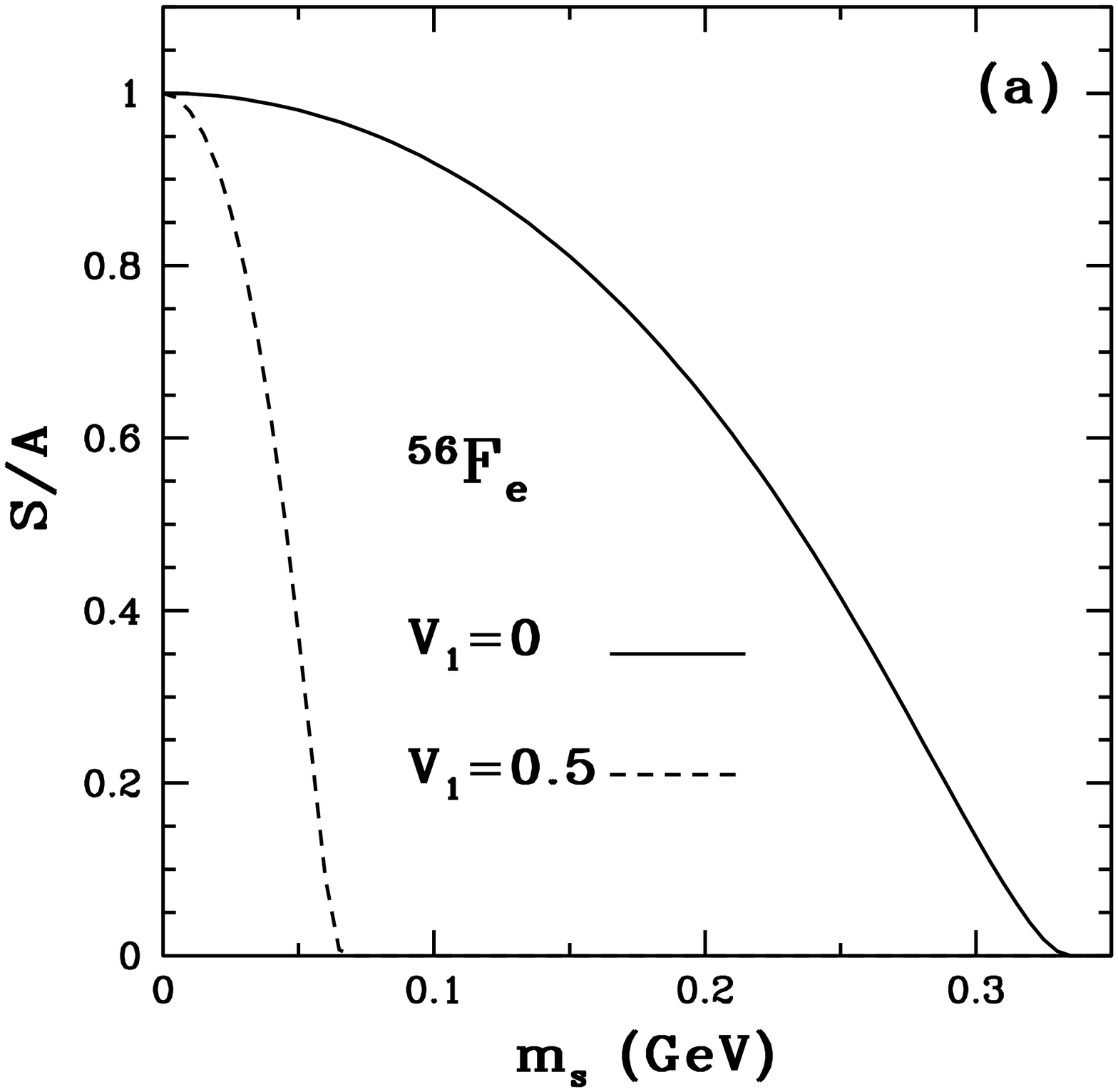,width=3.2truein,height=3.2truein}
\psfig{figure=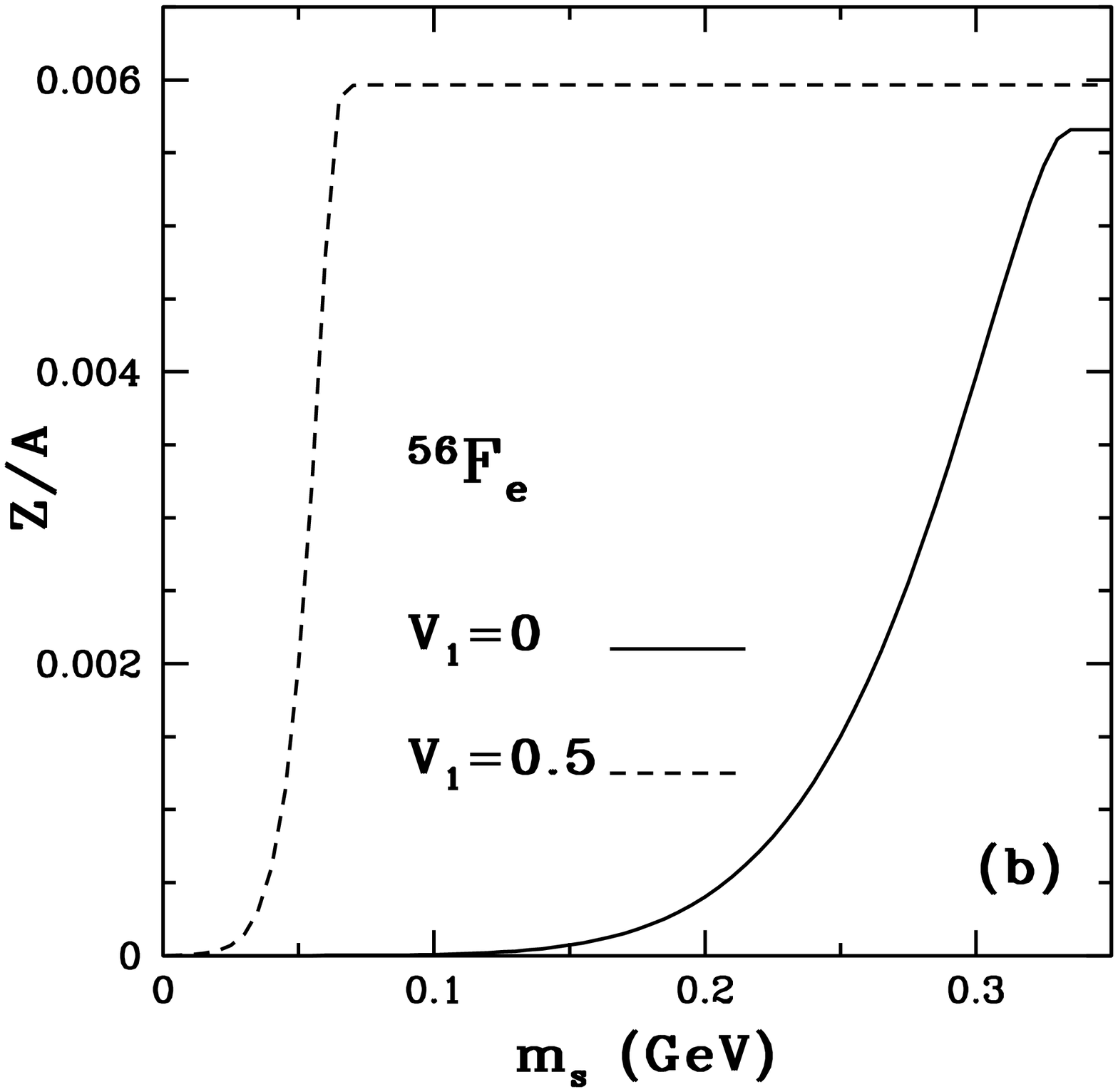,width=3.2truein,height=3.2truein}
\hskip .5in}
\caption{ 
}
\label{festr}
\end{figure*}

\begin{figure*}[th]
\centerline{
\psfig{figure=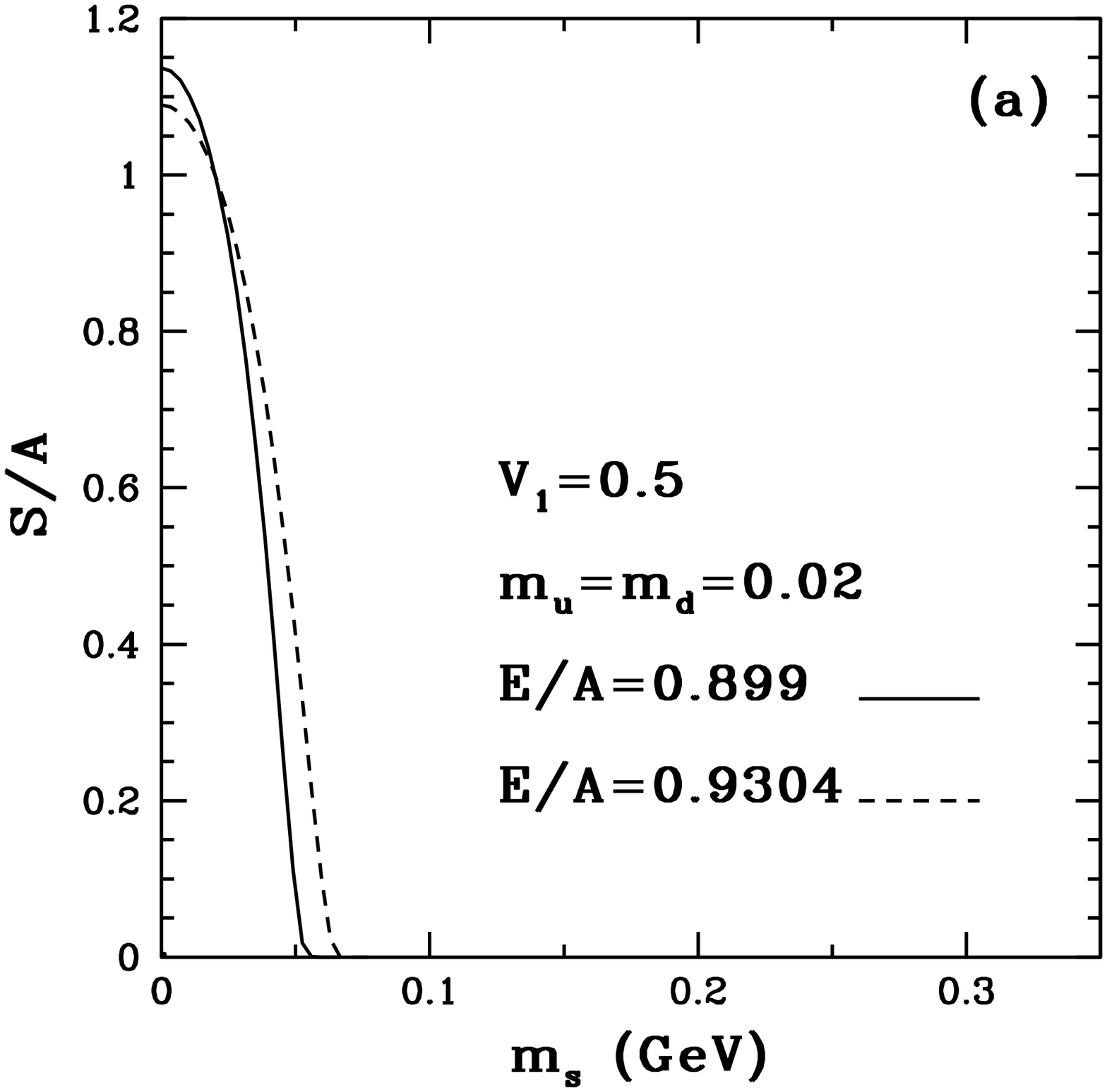,width=3.2truein,height=3.2truein}
\psfig{figure=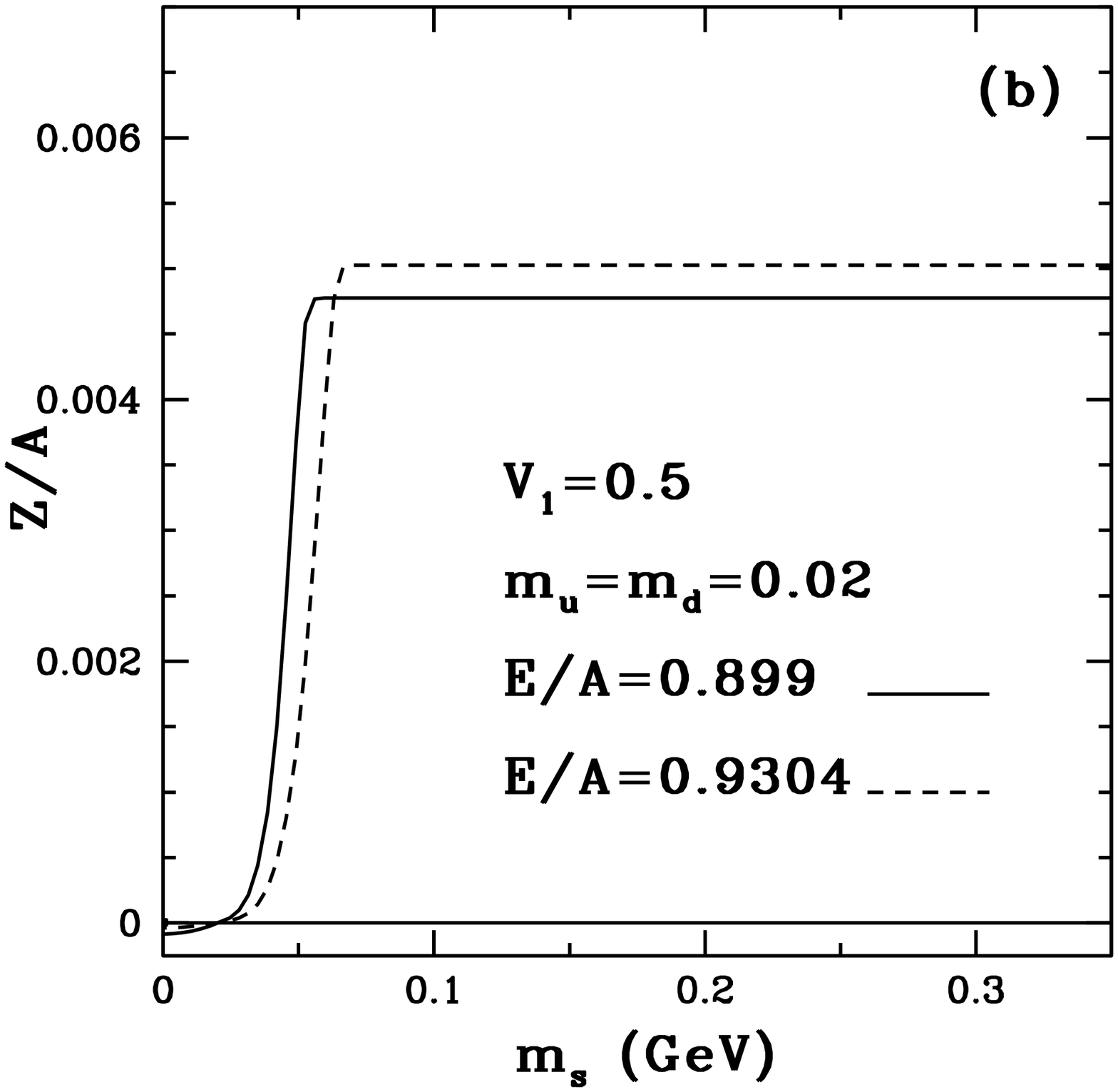,width=3.2truein,height=3.2truein}
\hskip .5in}
\caption{ 
}
\label{strqe}
\end{figure*}

 Taking into account the importance of experiments at RHIC and LHC, it is instructive at this  
point to consider finite temperature effects on the SQM properties.
 In order to check the influence of nonzero temperatures on the SQM stability at zero pressure, 
we have applied the same procedure employed for the $T=0$ case to study the behavior of the 
stability window shown in Fig. \ref{fewin}, but for $T\neq0$. 
 We have taken several values of $T$ up to 30 MeV for constant $V_1$ and for $V_1(T)$ parametrized in 
\cite{ST2} for $T\geq T_c\;$ as 
\bq
V_1(T)=\frac{0.175\;{\rm GeV}}{1.35\;T/T_c-1}\;,\;\;\;\;V_1(T_c)=0.5\;{\rm GeV}\;,
\lb{V1T}
\eq
for $T=T_c,\;2T_c,\;3T_c$ and some arbitrary values of $T_c$ along the phase diagram transition curve. 
 In both cases, the results are qualitatively analogous to those in Fig. 1 of \cite{KWW} and 
Fig. 3 of \cite{LuB}. 
 However, $V_1(T)$ is decreasing with the growth of $T$, so the shift of the stability window 
towards lower values of $G_2$ takes place as $T\rightarrow T_c$.
 For $V_1(T=T_c)$ the result is the same as for constant $V_1=0.5$ GeV at $T=0$ shown in panel 
(a) of Fig. \ref{fewin}. 

 Generally speaking, our results appear to be consistent with the fact that absolutely stable 
SQM has not been observed up to the present. 
 The experiment with STAR at RHIC has not confirmed the existence of SQM nor proved that it does 
not exist \cite{Sdw,STAR,MBl}. 
 The low values of $G_2$ with respect to the ones in \cite{ST1,ST2} would provide a possible 
explanation for the absence of absolutely stable SQM signature.
 However, before any conclusion towards the nonexistence of absolutely stable SQM,  we must 
have in mind that our theoretical results should be a consequence of the approximations 
contained in the development of the FCM nonperturbative EOS.
 FCM is a robust theoretical approach where the dynamics of confinement is one of the most 
important aspects of the model. 
 Therefore, we must be aware that the FCM nonperturbative EOS is presently developed in the 
so called Single Line Approximation, where the confinement dynamics include only single quarks 
and gluons interactions with the vacuum \cite{Si6}.

 On the other hand, in our calculations, $V_1$ and $G_2$ were taken as $\mu$-independent 
parameters.
  As pointed out in \cite{Bal,Bur}, the $\mu$-independence of $V_1$ should be a questionable 
assumption.
  Also, in the large density domain, important effects should be related to a possible density 
dependence of $G_2$ \cite{BCZ}.
 In our opinion, these aspects are very interesting possibilities to be considered. 
 However, this does not have been the scope of the present work.
 
\centerline{\bf ACKNOWLEDGMENTS} 

This work was done with the support provided by the Minist\'erio da Ci\^encia , Tecnologia 
e Inova\c c\~ao (MCTI).
  
%
%

\newpage

\centerline{\bf FIGURE CAPTION}

{\bf Fig. 1 -} 
 Contours of constant $E/A$ of strange quark matter at zero temperature 
and pressure for two different choices of the parameter $V_1$ 
($E/A$ and $V_1$ in GeV units). 
 The vertical dashed line is where the energy per baryon number of the 
two-flavor quark matter exceeds the one of $^{56}F_\ex$.
 Panels (a) and (c): strange quark mass as function of $\Delta\varepsilon_{\rm vac}$.
 Panels (b) and (d): strange quark mass as function of $G_2$. 
 The contours labeled (N) stand for the nucleon mass.
 The 0.93 contours correspond to 930.4 MeV energy per nucleon of $^{56}F_\ex$.

{\bf Fig. 2 -} 
 Panel (a): The strangeness per baryon as function of the strange-quark mass 
for different choices of $V_1$.
 Panel (b): As in panel (a), but for the baryon number density.
 Panel (c): As in panel (b), but for the hadronic electric charge per baryon.

{\bf Fig. 3 -} 
 Panel (a): Stability windows of SQM at zero temperature and pressure, 
bounded by the $E/A=0.93$ GeV contour of $\;^{56}F_\ex$ and its vertical line,   
for different choices of $V_1$ (in GeV units) labeling each contour.
 Panel (b): for each value of $V_1$ in panel (a), the respective locations of 
the $E/A$ contour (solid) and its vertical line (dashed) on the horizontal axis. 
 Panel (c): Baryon number densities at the energy per baryon of $\;^{56}F_\ex$ as 
function of the strange quark mass for different values of $V_1$.
 For $V_1=0.09$ GeV and $V_1=0.167$ GeV, we have $n_A=n_0$ at $m_s\sim0.3$ GeV and 
$m_s=0$, respectively, where $n_0=0.153\;{\rm fm}^{-3}$ is the nuclear saturation number density.

{\bf Fig. 4 -}
 Panel (a): Strangeness per baryon as function of the strange-quark mass 
for the given values of $V_1$ (in GeV units), all at $E/A=0.9304$ GeV 
of $\;^{56}F_\ex$.
 Panel (b): As in panel (a), but for the hadronic electric charge per baryon.
  
{\bf Fig. 5 -} 
 Panel (a): Strangeness per baryon as function of the strange-quark mass 
for the given nonzero u and d quark masses, $V_1$ and $E/A$ (all in GeV units). 
The energy per baryon $E/A=0.899$ GeV was taken for the purpose of comparison 
with the results in \cite{Far}. 
 Panel (b): As in panel (a), but for the hadronic electric charge per baryon.

\newpage

\newpage

\newpage

\newpage

\newpage

\end{document}